\documentclass[traditabstract]{aa}
\usepackage{epsfig}    
\usepackage{lscape}
\usepackage{natbib}
\bibpunct{(}{)}{;}{a}{}{,}  
\usepackage{graphics}
\usepackage{graphicx,graphics}
\usepackage{color} 
\usepackage{times}
\usepackage{amsmath}
\usepackage{amssymb}
\begin{document}

\title{Spiral galaxies with flat radial abundance gradients at large radii}

\author{
         L.~S.~Pilyugin\inst{\ref{ITPA},\ref{MAO}}            \and 
         G.~Tautvai\v{s}ien\.{e}\inst{\ref{ITPA}}     
}
\institute{Vilnius University, Faculty of Physics, Institute of Theoretical Physics and Astronomy,  Sauletekio av. 3, 10257, Vilnius, Lithuania \label{ITPA} 
\and
  Main Astronomical Observatory, National Academy of Sciences of Ukraine, 27 Akademika Zabolotnoho St, 03680, Kiev, Ukraine \label{MAO}
}

\abstract{
  We consider the oxygen abundance distributions for a sample of  massive (log($M_{\star}/M_{\sun}$) $\ga$ 10) spiral galaxies from the Mapping Nearby Galaxies at the Apache Point
  Observatory (MaNGA) survey  in which the radial abundance gradient flattens to a constant value outside of the outer break radius, $R_{\rm b,outer}$. The outer break radius
  can be considered as a dividing radius between the galaxy and the circumgalactic medium (CGM). The values of the $R_{\rm b,outer}$ range from $\sim 0.8\,R_{25}$ to $\sim1.45\,R_{25}$,
  where $R_{25}$ is the optical (isophotal) radius of the galaxy. The oxygen abundances in the CGM range from 12+log(O/H) $\sim 8.0$ to $\sim 8.5$. The oxygen abundance
  distribution in each of our galaxies also shows the inner break in the radial abundance profile at the radius $R_{\rm b,inner}$.   The metallicity gradient in the outer part of the galaxy
  ($R_{\rm b,inner} < R < R_{\rm b,outer}$) is steeper than in the inner part ($R < R_{\rm b,inner}$). The behaviour of the radial abundance distributions in these galaxies can be explained
  by assuming an interaction with (capture of the gas from) a small companion and adopting the model for the chemical evolution of galaxies with a radial gas flow. The interaction
  with a companion results in the mixing of gas and a flat metallicity gradient in the CGM. The capture of the gas from a companion increases the radial gas inflow rate and changes
  the slope of the radial abundance gradient in the outer part of the galaxy.
}


\titlerunning{outer and inner breaks in the metallicity gradients in spirals}
\authorrunning{Pilyugin and Tautvai{\v s}ien\.{e}}

\maketitle

\section{Introduction}

The discs of spiral galaxies have long been known to show negative radial abundance gradients in the sense that the abundance is higher at the centre and decreases with galactocentric
distance \citep{Searle1971,Smith1975}. The nebular abundance distributions within the optical (isophotal) radius of a galaxy $R_{25}$ (which is the galactocentric
distance of the isophote at a surface brightness of 25 mag$_{B}$ arcsec$^{-2}$, corrected for the galaxy inclination) were determined for many galaxies
by different authors \citep[][among many others]{VilaCostas1992,Zaritsky1994,Pilyugin2004,Pilyugin2007,Pilyugin2014,Pilyugin2019,Gusev2012,Sanchez2014,Ho2015,Zinchenko2015,Zinchenko2016,
SanchezMenguiano2016,SanchezMenguiano2018,Kreckel2019,Berg2020}. The general trend of the decreasing gas-phase oxygen abundance along the radius is well established.
\citet{Sanchez2014} found that all galaxies without clear evidence of an interaction present a common gradient in the oxygen abundance within the optical radius 
with a characteristic slope of $-0.16$~dex/$R_{25}$  and a dispersion of  0.12~dex/$R_{25}$.
It is also well established that  the radial abundance profile can show a  break within the optical radius of the galaxy, that is, the gradient in the slope of the radial abundance can
change \citep[e.g.][]{VilaCostas1992,Zaritsky1994,SanchezMenguiano2016,SanchezMenguiano2018,Pilyugin2024,Cardoso2025}. 

The abundance distributions beyond the optical radius of a galaxy were investigated less frequently. 
\citet{Bresolin2009} obtained spectra of 49 H\,{\sc ii} regions in the spiral galaxy NGC~5236 (M~83). The targeted H\,{\sc ii} regions span a range of galactocentric distances between
0.64 and 2.64 times the optical radius $R_{25}$, and 31 of them are located at $R > R_{25}$. The authors found that the oxygen abundances within the optical radius follow the
radial gradient obtained in other investigations, but an abrupt discontinuity in the radial oxygen abundance trend was detected near the optical radius of the disc. The outer abundance
trend flattens to an approximately constant value. 
\citet{Werk2011} carried out multi-slit optical spectroscopy and estimated  strong-line oxygen abundances of $\sim 100$ H\,{\sc ii} regions with projected galactocentric distances
ranging from 0.3 to 2.5 $R_{25}$ in 13 galaxies, including 8 massive ($\ga$10$^{10}M_{\sun}$) galaxies. 
They found that all these galaxies have flat radial oxygen abundance gradients from their central optical bodies to their outermost regions.
\citet{Bresolin2012} obtained spectra of 135 H\,{\sc ii} regions in the spiral galaxies NGC~1512 and NGC~3621 that span a range of galactocentric distances from 0.2 to 2~$R_{25}$.
The radial abundance gradient in these two galaxies flattens to a constant value outside the isophotal radius. The O/H abundance ratio in the outer disc is highly homogeneous, with
a scatter of only $\sim 0.06$~dex. 
\citet{Patterson2012} measured spectra of H\,{\sc ii} regions in NGC~3031 (M~81) to more than two optical radii $R_{25}$.
They found an overall oxygen abundance gradient in the entire radial range.
\citet{Grasha2022} studied in a spatially resolved H\,{\sc ii} region the gas-phase metallicity  maps of six local star-forming and face-on spiral galaxies from the  TYPHOON program.
They found that the metallicity gradient in the galaxy NGC~1566 is flat at a galactocentric distance greater than $\sim15$~kpc.

\citet{Sanchez2012b} pointed out a possible flattening of the radial gradients in the outer regions of spiral galaxies observed by the Calar Alto Legacy Integral Field Area survey
(CALIFA; \citealt{Sanchez2012a}). They later confirmed that the oxygen abundance of most of the 94 galaxies with H\,{\sc ii} regions detected beyond two effective disc radii becomes
flatter \citep{Sanchez2014}.  The authors noted that the flattening in the outer regions seems to be a universal property of disc galaxies, regardless of the mass, luminosity, morphology,
and presence of bars. \citet{SanchezMenguiano2016} measured the gas abundance profiles in a sample of 122 face-on spiral galaxies from the CALIFA survey. 
They found that the abundance gradient of most of the galaxies in the sample with reliable oxygen abundance values beyond two effective radii became flatter in
these outer regions. This flattening is not associated with any morphological parameter.
\citet{SanchezMenguiano2018} obtained the radial distribution of the oxygen abundances in a sample of 102 spiral galaxies observed with Very Large Telescope/Multi Unit Spectroscopic
Explorer (VLT/MUSE) . They found that 10 galaxies showed an outer flattening and 16 galaxies simultaneously showed an outer flattening and  an inner drop in the abundances. \citet{Cardoso2025} noted
that the nature of this ﬂattening in the external gradient is still debated because it is unclear whether this phenomenon is a common feature in galactic discs.

We consider the abundance distributions in a sample of  massive (log($M_{\star}/M_{\sun}$) $\ga$ 10) spiral galaxies from the Mapping Nearby Galaxies at the Apache Point
Observatory (MaNGA: \citealt{Bundy2015}) survey in which the radial abundance gradient flattens to a constant value at large galactocentric distances. Based on this, we examine the properties of the outer breaks
(transition to the region with a flat gradient) and ascertain the origin of this flattening.
We also wish to determine whether there is any other distinctive characteristic common to all these galaxies.
If this is the case, then this characteristic can also be related to the reason for the flat abundance distribution at large radii, that is, these characteristics are expected to originate from the same evolutionary pathways. 

The paper is organised in the following way. The data and a sample of selected galaxies are described in Sect.~2. In Sect.~3 we discuss the properties of the investigated galaxies. Section~4 contains a brief summary.

\begin{table*}
\caption[]{\label{table:general}
   Characteristics\tablefootmark{a} of the MaNGA galaxies
}
\begin{center}
\begin{tabular}{cccccccccccc} \hline \hline
Galaxy                  &
$d$                     &
log\,$M_{\star}$          &
$R_{25}$                 &
$R_{\rm b,inner}$          &
$R_{\rm b,outer}$          &
(O/H)$_{\rm CGM}$         &
log\,SFR                &
$A$                     &
$R_{\rm A}$               &
environment             & 
morphology             \\
ID                     &
Mpc                    &
$M_{\sun}$              &
Kpc                    &
$R/R_{25}$              &
$R/R_{25}$              &
12+log(O/H)            &
$M_{\sun}$/year         &
                       &
                       &
                       &
                       \\
\hline
(1)                    &
(2)                    &
(3)                    &
(4)                    &
(5)                    &
(6)                    &
(7)                    &
(8)                    &
(9)                    & 
(10)                   & 
(11)                   & 
(12)                   \\
\hline
  7960 12704  &  108.4  &  10.452  &  11.30  &   0.500  &   0.850  &    8.335  &    0.455   &   0.248  &    0.127  & multu    & Sc    \\ 
  8091 12703  &  333.1  &  11.003  &  20.27  &   0.850  &   0.950  &    8.369  &    0.707   &   0.294  &    0.140  &          & S     \\ 
  8135 12703  &  225.9  &  10.668  &  13.36  &   0.790  &   1.153  &    8.358  &   -0.061   &   0.133  &    0.067  & isolated & Sbc   \\ 
  8247 12703  &  238.2  &  10.588  &  14.95  &   0.443  &   1.054  &    8.404  &    0.419   &   0.247  &    0.116  & isolated & Sc    \\ 
  8262 12704  &  220.2  &  10.518  &  16.28  &   0.824  &   0.997  &    8.194  &   -0.009   &   0.167  &    0.095  & multu    & Sc    \\ 
  8443 12703  &  260.1  &  11.390  &  27.87  &   0.606  &   0.997  &    8.425  &    0.216   &   0.180  &    0.085  & multu    & Sb    \\ 
  8595 12702  &  210.9  &  10.460  &  14.42  &   0.945  &   1.106  &    8.172  &   -0.008   &   0.242  &    0.120  & isolated & SBb   \\ 
  8613 12702  &  146.9  &  10.910  &  16.42  &   1.124  &   1.240  &    8.314  &   -0.194   &   0.138  &    0.068  & multu    & SBc   \\ 
  9000 12702  &  126.0  &  10.259  &  13.29  &   0.720  &   1.474  &    8.137  &   -0.130   &   0.214  &    0.113  & multu    & SBbc  \\ 
  9024 12701  &  397.4  &  11.466  &  28.32  &   1.017  &   1.183  &    8.434  &   -0.134   &   0.113  &    0.060  & multu    & Sb    \\ 
  9025 12705  &  219.1  &  10.428  &  14.66  &   0.745  &   0.972  &    8.298  &   -0.253   &   0.145  &    0.077  & isolated & Sc    \\ 
  9485 12705  &  144.8  &  11.125  &  15.76  &   0.481  &   0.876  &    8.519  &    0.476   &   0.145  &    0.076  & isolated & SBa   \\ 
  9492 12703  &  207.7  &  10.390  &  15.20  &   0.622  &   1.026  &    8.262  &    0.055   &   0.129  &    0.073  & isolated & Sc    \\ 
  9506 12704  &  218.9  &  10.732  &  14.64  &   0.999  &   1.100  &    8.219  &   -0.047   &   0.120  &    0.064  & isolated & Sbc   \\ 
  9879 12704  &  252.9  &  10.833  &  13.36  &   1.150  &   1.150  &    8.519  &    0.083   &   0.164  &    0.091  & multu    & SBb   \\ 
 10842 12703  &  328.8  &  11.215  &  19.13  &   0.948  &   1.176  &    8.385  &    0.272   &   0.109  &    0.058  & multu    & S0-a  \\ 
 10844 12705  &  283.8  &  11.062  &  14.24  &   0.951  &   1.174  &    8.309  &    0.794   &   0.294  &    0.129  & isolated & Sc    \\ 
 11945 12705  &  239.5  &  10.718  &  16.43  &   0.853  &   0.986  &    8.388  &    0.247   &   0.189  &    0.094  & multu    & Sbc   \\ 
 11946 12702  &  286.2  &  11.151  &  18.25  &   0.620  &   1.042  &    8.450  &    0.303   &   0.227  &    0.121  & multu    & S     \\ 
 12094 12705  &  228.7  &  10.323  &  10.75  &   1.001  &   1.001  &    8.169  &    0.712   &   0.277  &    0.133  &          & S     \\ 
                    \hline
\end{tabular}\\
\end{center}
\tablefoottext{a}{
  (1) MaNGA identificator; 
  (2) distance to the galaxy $d$; 
  (3) stellar mass $M_{\star}$; 
  (4) optical (or isophotal) radius $R_{25}$; 
  (5) radial position of the inner break, $R_{\rm b,inner}$, in the radial abundance distribution; 
  (6) radial position of the outer break, $R_{\rm b,outer}$;    
  (7) oxygen abundances of the circumgalactic medium, (O/H)$_{\rm CGM}$; 
  (8) current star formation rate; 
  (9) asymmetry parameter, $A$, which quantifies the  asymmetry of a light distribution across the galaxy;  
  (10) asymmetry index $R_{A}$, which specifies the distribution of residual fluxes  after the model flux is subtracted;   
  (11) environment;
  (12) morphological type.
} 
\end{table*}

\section{Data and mapping of the galaxy properties}

\subsection{Data}

Our investigation was based on galaxies from the MaNGA survey \citep{Bundy2015}. 
The completed observations of MaNGA galaxies are reported in Data Release 17 \citep{Abdurrouf2022}, where the data products were revised for all the observations that were previously
released in DR15 and before (e.g. the flux calibration was updated). The emission line parameters of the spaxel spectra of galaxies are available from the MaNGA Data Analysis
Pipeline (DAP) measurements. We derived the characteristics of a sample of galaxies using the last version of DAP measurements
manga-n-n-MAPS-SPX-MILESHC-MASTARSSP.fits.gz\footnote{https://data.sdss.org/sas/dr17/manga/spectro/analysis/v3\_1\_1/3.1.0/SPX-MILESHC-MASTARSSP/} for the spectral measurements and
the Data Reduction Pipeline (DRP) measurements manga-n-n-LOGCUBE.fits.gz\footnote{https://dr17.sdss.org/sas/dr17/manga/spectro/redux/v3\_1\_1/} for the photometric data.
\citet{Sanchez2022} derived different characteristics of the MaNGA galaxies by ﬁtting the spaxel spectrum with a combination of spectra of simple stellar populations (SSP) of different
ages and metallicities. These results are reported in their catalogue\footnote{https://data.sdss.org/sas/dr17/manga/spectro/pipe3d/v3\_1\_1/3.1.1/}.

We considered a sample of the discy galaxies for which the curves of isovelocities in the measured line-of-sight velocity fields look like a set of parabola-like curves
(hourglass-like picture of the rotation disc). This condition grants us the opportunity to derive the geometric parameters of a galaxy, which are necessary to determine the
galactocentric distances of individual spaxels and to construct radial distributions of the characteristics across the galaxy. This criterion also rejects 
interacting and merging galaxies when the interaction distorts the line-of-sight velocity field to such an extent that the determination of the geometrical angles and rotation curve
is impossible. We also excluded galaxies with an inclination angle larger than $\sim 70\degr$ because the fit of the H$\alpha$ velocity field in these galaxies can produce
unrealistic values of the inclination angle \citep{Epinat2008}, and as a result, the obtained galactocentric distances of the spaxels can involve large uncertainties.
We considered galaxies with a large number of spaxels in the galaxy image, that is, galaxies mapped with 91 and 127 fibre IFU, covering 27$\farcs$5 and 32$\farcs$5 on the sky. 
We only chose galaxies for which the spaxels with measured emission lines were well distributed across the galactic discs and covered more than $\sim 0.8~R_{25}$. Thus, we considered
430 MaNGA galaxies.

We took distances to the galaxies from the NASA/IPAC Extragalactic Database ({\sc ned})\footnote{The NASA/IPAC Extragalactic Database
({\sc ned}) is operated by the Jet Propulsion Laboratory, California Institute of Technology, under contract with the National Aeronautics and Space Administration.
{\tt http://ned.ipac.caltech.edu/}}.  The {\sc ned} distances use flow corrections for Virgo, the Great Attractor, and Shapley Supercluster infall (adopting a
cosmological model with $H_{0} = 73$ km/s/Mpc, $\Omega_{m} = 0.27$, and $\Omega_{\Lambda} = 0.73$).
We adopted the spectroscopic stellar masses from the Sloan Digital Sky Survey (SDSS) and BOSS \citep[BOSS stands for the Baryon Oscillation Spectroscopic Survey in SDSS-III, see][]{Dawson2013}.
The spectroscopic masses were taken from the table {\sc stellarMassPCAWiscBC03} and were determined using the Wisconsin method \citep{Chen2012} with the stellar population synthesis models
from \citet{Bruzual2003}.

\subsection{Determining the geometrical parameters and the galaxy size}

The geometrical parameters of the galaxy (the coordinates of the centre, the position angle of the major axis, the inclination angle, and the rotation curve) were derived through the best
fit to the observed line-of-sight gas velocity field (obtained from the measured wavelength of the emission of the H$\alpha$ line in the spaxel spectra) following \citet{Pilyugin2019}. 
The galactocentric distances of the spaxels determined with these geometrical parameters are used below to construct the radial distributions of the oxygen abundances
and other characteristics across the disc.

We specified the size of the galaxy by the isophotal (or optical) radius $R_{25}$, which is the galactocentric distance of the isophote at a surface brightness of 25 mag$_{B}$ arcsec$^{-2}$,
corrected for the galaxy inclination. The photometric profile of the galaxy was obtained in the following way. The measurements in the SDSS filters $g$ and $r$ for each spaxel were
converted into $B$-band magnitudes following \citet{Pilyugin2018}. The radial surface brightness distribution was constructed using the galactocentric distances of the spaxels
determined with the coordinates of the centre, the position angle of the major axis, and the inclination angle obtained from the analysis of the observed gas velocity field.
The radial surface brightness distribution was fitted within the optical radius by a broken exponential profile for the disc and by a general Sérsic profile for the bulge. 
The value of the isophotal radius $R_{25}$ was estimated using the fit corrected for the galaxy inclination. We used a simple correction for the inclination (factor cos\,$i$) for the
surface brightness of the disc component. The bulge was assumed to be spherical,  and its surface brightness was not corrected for inclination. Since we are mainly interested in the
surface brightness of the outer galaxy regions (determination of the optical radius), this approach is justified even when the bulge is not spherical.

\subsection{Determining the oxygen abundances}

\begin{figure*}
\begin{center}
\resizebox{1.00\hsize}{!}{\includegraphics[angle=000]{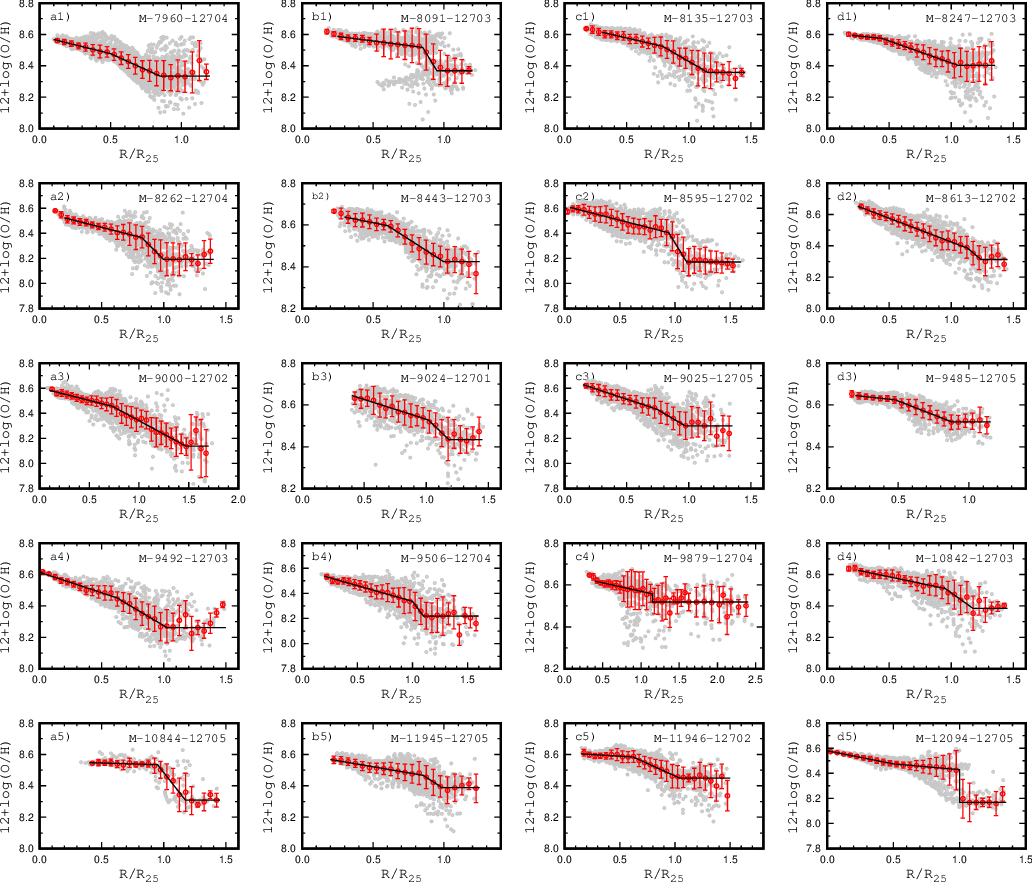}}
\caption{Radial oxygen abundance distributions for our sample of the MaNGA galaxies.
Each panel shows the radial oxygen abundance distribution for individual spaxels (grey points)  and binned (0.05$R_{25}$) values (red circles). The line denotes the adopted O/H -- $R$ relation.  
}
\label{figure:r-oh-manga}
\end{center}
\end{figure*}

The measured emission line fluxes in the spaxel spectra were corrected for the interstellar reddening using the reddening law of \citet{Cardelli1989} with $R_{V}$ = 3.1. 
The logarithmic extinction at H$\beta$ was estimated through a comparison of the measured and theoretical $F_{{\rm}H\alpha}/F_{{\rm H}\beta}$ ratios, where the theoretical
value of the line ratio of 2.87 was taken from \citet{Osterbrock2006},   assuming case $B$ recombination, an electron temperature of 10$^{4}$K, and an electron density of 100 cm$^{-3}$.
  When the measured value of the ratio $F_{{\rm}H\alpha}/F_{{\rm H}\beta}$ was lower than the theoretical value, the reddening was adopted to be zero.

The intensities of strong emission lines are usually used to separate different types of spaxel spectra according to their main  excitation mechanism (i.e., the
star forming (SF) and the active galactic nucleus (AGN) spectra). 
A widely used spectral classification of emission-line spectra is the [O\,{\sc iii}]$\lambda$5007/H$\beta$ versus \ [N\,{\sc ii}]$\lambda$6584/H$\alpha$ diagnostic diagram 
suggested by \citet{Baldwin1981}  (BPT classification diagram). 
\citet{Kauffmann2003}  found an empirical demarcation line between the star-forming and AGN-like spectra in the BPT diagram.  
This demarcation line  can be interpreted as an upper limit of pure star-forming  spectra. The spectra located to the left (below) of the demarcation line of \citet{Kauffmann2003}
are referred to as the H\,{\sc ii} region-like (or the SF-like) spectra. 
\citet{Kewley2001} determined a theoretical demarcation line between the star-forming and AGN spectra in the BPT diagram.
This demarcation line is a theoretical upper limit for starburst models in the diagnostic diagram. 
The spectra located to the right (above) of the demarcation line of \citet{Kewley2001} are referred to as AGN-like spectra. The spectra located between the demarcation lines of
\citet{Kauffmann2003} and \citet{Kewley2001}
are referred to as intermediate (INT) spectra. In the literature, these spectra are also referred  to as composite or 
transition spectra \citep[e.g.][]{Davies2014,Pons2014,Pons2016}. 

The credibility of the classification of the ionising source of the region using only the BPT diagram has been questioned, however
\citep[e.g.][]{Sanchez2014,Lacerda2018,D'Agostino2019,Sanchez2020,Sanchez2021,Sanchez2024,Sanchez2025}. 
It was suggested to use the equivalent width of the emission H$\alpha$ line, EW$_{{\rm H}\alpha}$, \citep{CidFernandes2010,CidFernandes2011,Sanchez2014,Lacerda2018,Sanchez2021,Sanchez2024}
and the gas velocity dispersion, $\sigma_{{\rm H}\alpha}$,  \citep{D'Agostino2019,Johnston2023,Sanchez2024} as diagnostic indicators in addition to the emission-line ratios, which are at the
base of the BPT diagnostic diagram.  \citet{Sanchez2024}  have proposed a method that explores the location of different ionising sources in a diagram, which compares the equivalent width
of the emission  H$\alpha$ line, EW$_{{\rm H}\alpha}$, and the gas velocity dispersion, $\sigma_{{\rm H}\alpha}$, (WHaD diagram). They defined different areas in which the ionising source could
be classified as (1) SF, ionisation due to young-massive OB stars, related to recent star formation activity; (2) sAGNs and wAGNs, ionisation due to strong (weak) AGNs, and other sources
of ionisations like high velocity shocks; and (3) Ret, ionisation due to hot, old low-mass evolved stars (post-AGBs), associated with retired regions within galaxies (in which there is
no star formation). They classified sources with a EW$_{{\rm H}\alpha} >$ 6 {\AA} and  $\sigma_{{\rm H}\alpha} <~$57~km\,s$^{-1}$ as SF. 
We estimated the oxygen abundances using the R calibration from \citet{Pilyugin2016} in the spaxels, where the spectra are the H\,{\sc ii} region-like spectra according
to the BPT and the WHaD classification diagrams simultaneously.

\subsection{Sample of galaxies with a flat abundance distribution in the circumgalactic medium}

We derived the radial abundance distribution not based on the abundances in individual spaxels, but on the median values of the abundances in bins of 0.05~dex in $R/R_{25}$ (red points in
Fig.~\ref{figure:r-oh-manga}). The radial oxygen abundance distribution was approximated by the broken linear relation. In the general case, we considered two breaks in the radial abundance
distribution (Fig.~\ref{figure:r-oh-manga}). The inner break, $R_{\rm b,inner}$, divides the inner and outer parts of the disc when the change in the gradient slope takes place within the disc.
The outer break, $R_{\rm b,outer}$, divides the disc of the galaxy and the circumgalactic medium (CGM).

We searched for galaxies with an outer break in the radial metallicity profile. The radial metallicity profile should be measured up to at least $\sim$1.2$R_{25}$
(see below). Unfortunately, the radial extension of the area with the measured abundance was not large enough for most of the MaNGA galaxies.  
We selected galaxies for which the oxygen abundances beyond the outer break were estimated in at least five bins. This request meant that we were able to determine a more or less reliable value
of $R_{\rm b,outer}$ and to estimate the abundance beyond the outer break (in the circumgalactic medium). We found 20 MaNGA galaxies that satisfied this criterion.
Is should be noted that we found no galaxies with an undisputable lack of the outer break for galaxies whose radial metallicity profile was measured up to more $\sim$1.2$R_{25}$. 
  The morphological types of the selected galaxies extended from {\sl So-a} to {\sl Sc}  according to the HyperLeda\footnote{http://leda.univ-lyon1.fr/} database \citep{Makarov2014},
  and five of them are classified as barred galaxies.
The radial abundance distributions for the selected galaxies are shown in  Fig.~\ref{figure:r-oh-manga}, and the general characteristics and parameters of the radial abundance distribution
are reported in Table~\ref{table:general}.  These galaxies are discussed in the next section. 

The validity of the obtained broken metallicity gradients can be argued in the following way. First, the credibility of the R calibration-based abundances was tested in our previous papers
and by other investigators. In particular, \citet{Croxall2016} compared R calibration-based abundances and their T$_{e}$-based abundances for NGC~5457. They found that  the
R calibration successfully reproduces the abundance gradient within the  1$\sigma$ errors throughout the whole metallicity range of around an order of magnitude. Second, we presented a sample of
29 galaxies in which the  R calibration-based oxygen abundance gradient was approximated by a single linear relation throughout the whole disc, with a mean value of the deviation from the relation
lower than $\sim$0.01 dex for the binned oxygen abundance \citep{Pilyugin2024}. This is evidence that the R calibration does not produce the false break in metallicity gradient, and
the obtained breaks in the metallicity gradients in our target galaxies are therefore real.

\section{Discussion}

\subsection{Properties of the abundance distributions}

\begin{figure}
\begin{center}
\resizebox{1.00\hsize}{!}{\includegraphics[angle=000]{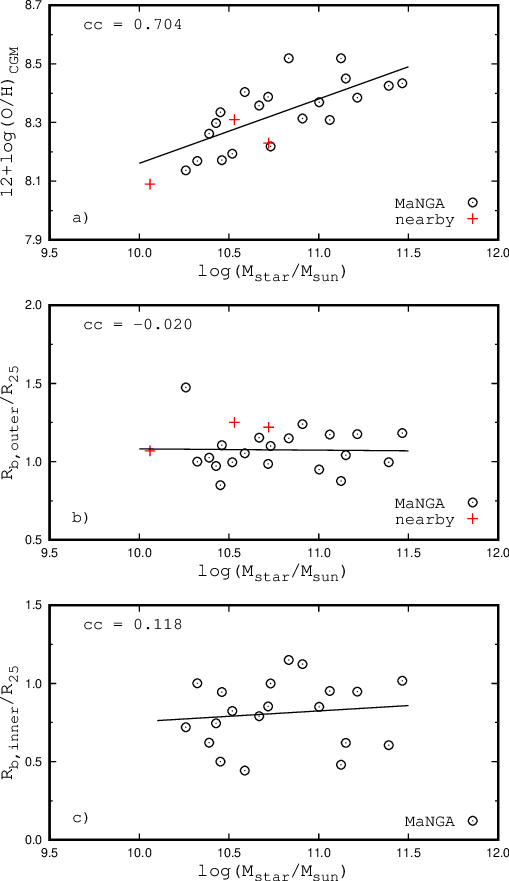}}
\caption{Characteristics of the radial oxygen abundance distributions for our sample of galaxies.
  The circles designate the oxygen abundance in the circumgalactic medium ({\em panel {\bf a}}),
  the radial positions of the outer break  ({\em panel {\bf b}}), and the radial positions of the inner break  ({\em panel {\bf c}}) 
  as a function of the stellar mass of MaNGA galaxies.
  The line is the linear best fit to these data points.
  The plus signs mark the nearby galaxies. 
}
\label{figure:rb-m}
\end{center}
\end{figure}

\begin{figure}
\begin{center}
\resizebox{1.00\hsize}{!}{\includegraphics[angle=000]{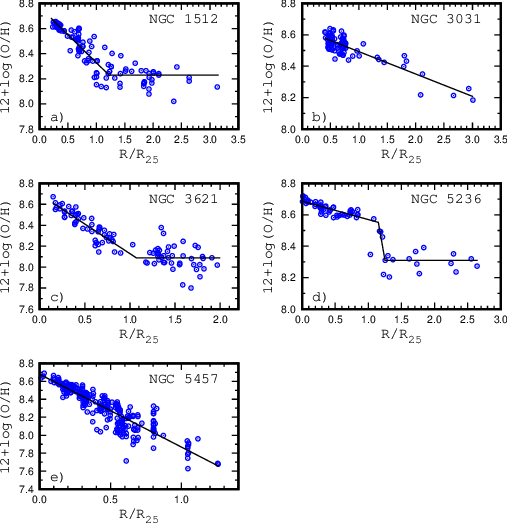}}
\caption{Radial oxygen abundance distributions for nearby galaxies.
The circles in each panel show the oxygen abundances for individual  H\,{\sc ii} regions as a function of radius. The line denotes the adopted O/H -- $R$ relation.  
}
\label{figure:r-oh-nearby}
\end{center}
\end{figure}

First, we examined whether the galaxies with the flat abundance of the CGM might be linked to their present-day environments. Eighteen out of 20 galaxies in our sample are listed in a catalogue of
galaxy groups and clusters \citep{Tempel2018}.  According to this catalogue, 10 galaxies in our sample are members of galaxy pairs or groups, and 8 galaxies  are classified as
isolated galaxies (Table~\ref{table:general}). This suggests that the flat abundance of the CGM  is independent of the present-day galaxy environment. 
Next,  we considered the evolutionary status of our galaxies, that is, we examined their positions in the diagram of stellar mass versus star formation rate (SFR).
The global star formation rate in a galaxy was estimated from the H$\alpha$ luminosity of a galaxy using the calibration relation of \citet{Kennicutt1998}, reduced by \citet{Brinchmann2004}
for the Kroupa initial mass function \citep{Kroupa2001}.  
The  H$\alpha$ luminosity of a galaxy was obtained as a sum of the H$\alpha$ luminosities of the spaxels with  H\,{\sc ii}-region-like spectra within the optical
radius.  We found that the SFRs in our galaxies agree with the SFR -- $M_{\star}$ relation for the present-day epoch ($t$ = 13.6 Gyr) from \citet{Speagle2014}. 

Panel (a) in Fig.~\ref{figure:rb-m} shows the oxygen abundance in the circumgalactic medium, (O/H)$_{\rm CGM}$, as a function of the stellar mass of the galaxy $M_{\star}$.
The values of (O/H)$_{\rm CGM}$ range from 12 + log(O/H) $\sim$8.0 to  $\sim$8.5. 
There is a correlation (correlation coefficient is equal to 0.704) between (O/H)$_{\rm CGM}$ and  $M_{\star}$. 
This is in line with the results of \citet{SanchezMenguiano2016} that the abundance value of the flattening depends on the mass of galaxies. 

Panel (b) in Fig.~\ref{figure:rb-m} shows the radial position of the outer break, $R_{\rm b,outer}$, as a function of the stellar mass of the galaxy.
The $R_{\rm b,outer}$ does not correlate with $M_{\star}$; the correlation coefficient is $-0.020$. The mean value for our sample is  $R_{\rm b,outer}/R_{25}$ = 1.08 with
a scatter of 0.14.
\citet{Bresolin2012} obtained for several nearby galaxies that the radial abundance gradient flattens to a constant value  near the optical radius of the disc. 
\citet{SanchezMenguiano2018} found a much wider distribution for the outer breaks. The flattening occurred there between 0.3 and 2.8~$R_{\rm e}$. 
The authors noted that although the distribution peaks at $\sim$1.5$R_{\rm e}$ (the mean value of the radial positions is 1.47$R_{\rm e}$ and the standard deviation 0.60), the wide
range of radii covered by it prevents the defininition of a characteristic location of the outer flattening in the abundances.

Panel (c) in Fig.~\ref{figure:rb-m} shows the radial position of the inner break, $R_{\rm b,inner}$, as a function of the stellar mass of the galaxy.
The values of $R_{\rm b,inner}$ for our sample of galaxies range from $\sim 0.45$ to $\sim 1.15$ of the optical radius $R_{25}$. 
There is no appreciable correlation between $R_{\rm b,inner}$ and $M_{\star}$; the correlation coefficient is 0.118.
\citet{SanchezMenguiano2018} have found that the distribution of the radial positions of the inner break is quite narrow, and a pronounced peak is centred at $\sim 0.5$ of the effective
radius $R_{\rm e}$ (the mean value of the radial positions is 0.54~$R_{\rm e}$ and the standard deviation 0.20), suggesting that the position of the inner drop is very similar in
all galaxies with this feature.  
\citet{Cardoso2025} have found that the average value for the position of the inner drop is at $R_{\rm b,inner}$ = 0.84$\pm$0.26 $R_{\rm e}$ and that  
the inner drop in massive galaxies lies closer to the galaxy centre. Thus, there is a significant difference between the inner break radii obtained here and by 
\citet{SanchezMenguiano2018} and \citet{Cardoso2025}.

It was noted above that the radial oxygen abundance profiles in several nearby galaxies were measured to more than two optical radii $R_{25}$.
We compared the radial oxygen abundance profiles in these nearby galaxies and in our MaNGA galaxies.  Fig.~\ref{figure:r-oh-nearby} shows the radial abundance distributions in
the nearby galaxies.  The oxygen abundances in the H\,{\sc ii} regions were estimated using the R calibration from \citet{Pilyugin2016} with the line fluxes from the sources
reported by  \citet{Pilyugin2023}. The geometrical  characteristics of the galaxies (the inclination angle of a galaxy, the position angle of the major
axis, the distance, and the angular and physical optical radii) used in the determinations of the galactocentric distances to the H\,{\sc ii} regions were also taken from  \citet{Pilyugin2023}.
The outer break can be seen in three nearby galaxies NGC~1512, NGC~3621, and NGC~5236 (panels (a), (c), and (d) in Fig.~\ref{figure:r-oh-nearby}, respectively). 
The only outer break was considered in the fitting of the radial oxygen abundance distribution because the radial position of the inner break radius (if exists) cannot be reliably established
on the basis of available measurements. The positions of these nearby galaxies in the (O/H)$_{\rm CGM}$ -- $M_{\star}$ and the $R_{\rm b,outer}$ -- $M_{\star}$ diagrams
(panels (a) and (b) in Fig.~\ref{figure:rb-m}, respectively) follow the bands outlined by our sample of the MaNGA galaxies.

The nearby galaxy NGC~3031 (M~81) shows an overall oxygen abundance gradient in the entire radial range up to around $\sim$3$R_{25}$ without an evident outer break
(panel (b) in Fig.~\ref{figure:r-oh-nearby}). \citet{Patterson2012} noted, however, that more deep spectroscopic data for H\,{\sc ii} regions are needed to solidly establish 
whether this galaxy shows a broken abundance profile with a flat outer gradient.
We also show the abundance profile for the nearby galaxy NGC~5457 (M~101) in panel (e) of  Fig.~\ref{figure:r-oh-nearby}, although the measurements of H\,{\sc ii} regions in this galaxy 
are only available up to $\sim$1.25$R_{25}$. This galaxy demonstrates that the oxygen abundance can decrease monotonically (without a break) along the radius by  around an order of
metallicity magnitude and that the abundance near the optical radius in a large spiral galaxy can be low, 12 + log(O/H) $\sim$ 7.8.

We found no analogues to the nearby galaxies NGC~3031 (with an overall oxygen abundance gradient in the entire radial range up to at least two optical radii)
and to the nearby galaxy NGC~5457 (with unprecedently large radial variation in the  oxygen abundance across the disc) among the MaNGA galaxies.  The lack of the analogues to NGC~3031
can be attributed to the fact that the abundances at large radii are not measured in the majority of the MaNGA galaxies. The lack of NGC~5457 analogues among hundreds of the MaNGA
galaxies is amazing, however. Even some nearby galaxies apparently might have a unique (or at least a rare way of) evolution.

\subsection{Origin of the flat abundance distribution in the CGM}

\begin{figure*}
\begin{center}
\resizebox{0.90\hsize}{!}{\includegraphics[angle=000]{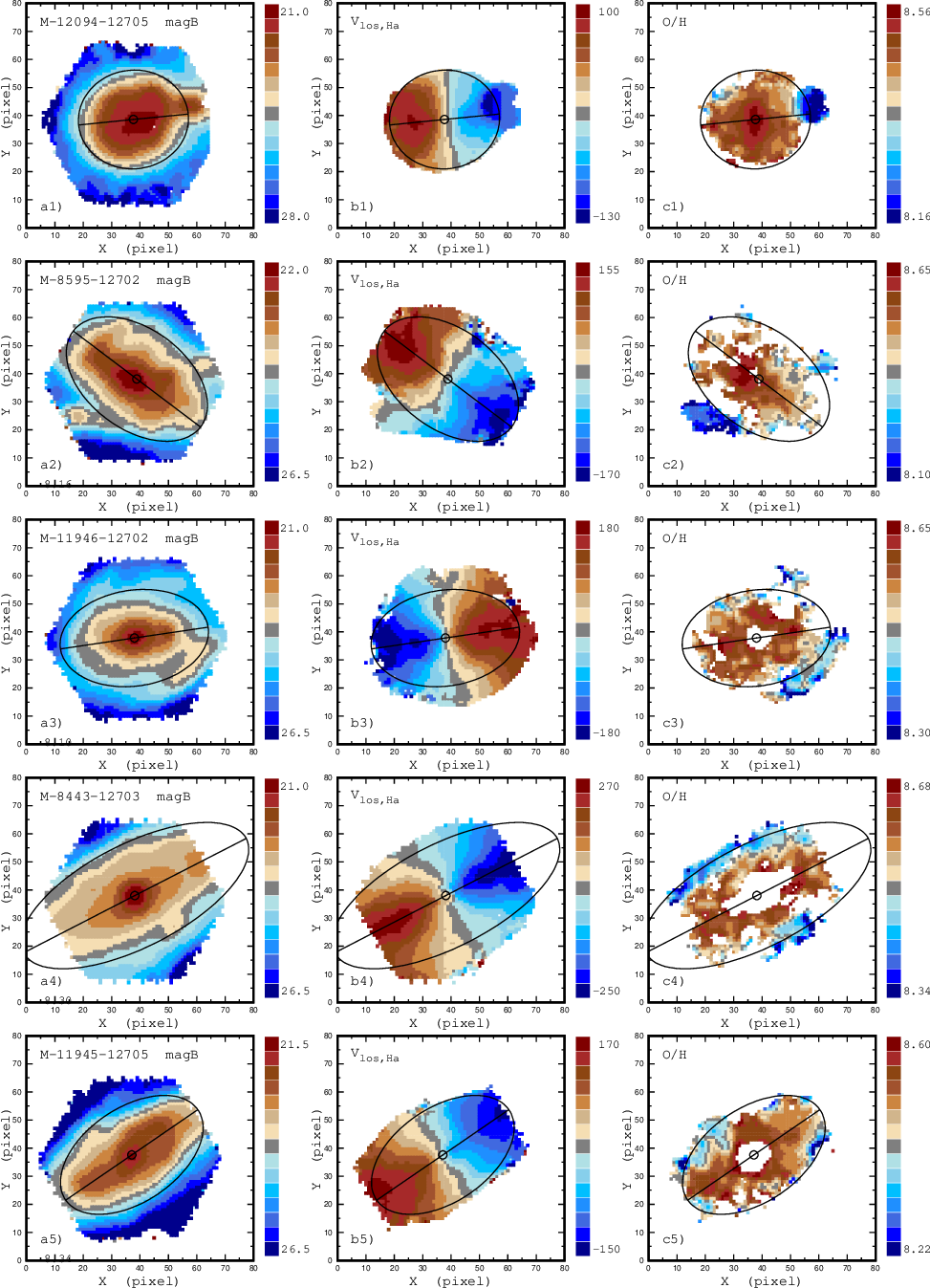}}
\caption{Characteristics maps for five galaxies.
{\em Panels of the left column} {\bf a:} Observed surface brightness distribution in the galaxy image in sky coordinates (pixels). The surface brightness value is colour-coded.
The circle shows the kinematic centre of the galaxy, the line indicates the position of the major kinematic axis of the galaxy, and the  ellipse is its optical radius.
{\em Panels of the middle column} {\bf b:} Line-of-sight velocity field. 
{\em Panels of the right column} {\bf c:} Distribution of O/H in the galaxy image.
}
\label{figure:sequence}
\end{center}
\end{figure*}

\begin{figure}
\begin{center}
\resizebox{1.00\hsize}{!}{\includegraphics[angle=000]{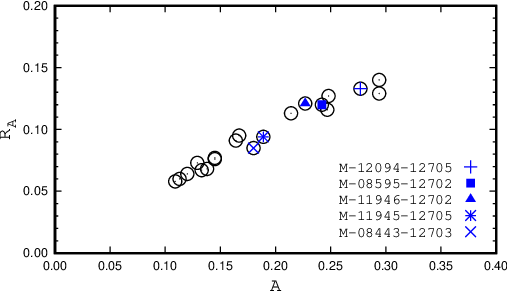}}
\caption{Asymmetry of a light distribution in the galaxies of our sample. 
  The asymmetry index of residual fluxes after the model flux is subtracted, $R_{A}$, vs. the asymmetry parameter in a light distribution for the galaxy, $A$.
  The circles denote all the galaxies, and the galaxies shown in Fig.~\ref{figure:sequence} are marked by symbols shown in the legend.  
}
\label{figure:a-ra}
\end{center}
\end{figure}

The flat abundance gradient in the circumgalactic medium can originate in different ways. 
It was noted that a levelling out of the star formation efficiency, defined as the ratio of the surface density of the star formation rate to the gas surface density, 
above and beyond the isophotal radius can explain the flattening of the chemical gradient observed in the circumgalactic medium \citep{Bresolin2012,Esteban2013}. 
However, the oxygen abundances in the circumgalactic medium are rather high, between 12 + log(O/H) $\sim 8.0$ and  12 + log(O/H) $\sim 8.5$  (Table~\ref{table:general}).  
\citet{Bresolin2012} noted that although a fraction of the metal amounts found in the circumgalactic medium are expected to be produced in situ, another
considerable fraction must have another origin. 

The flat abundance gradient would also be observed if the circumgalactic medium were well mixed. Models predict that the interaction of galaxies results in gas mixing, and this consequently
produces a flat abundance gradient \citep[e.g.][]{Rupke2010}. \citet{Werk2011} concluded that there is efficient metal mixing out to large galactocentric radii in their sample of galaxies,
with flat radial oxygen abundance gradients from their central optical bodies to their outermost regions, facilitated by galaxy interactions. Six out of eight massive galaxies from their 
sample are indeed classified as minor merges or interacting galaxies. \citet{Werk2011} noted that the interactions might be the primary driver of the metal mixing in these galaxies, and the
mechanism by which it acts is independent of the interaction details.

The contribution of gas captured from a dwarf galaxy by the circumgalactic medium can also explain the rather high oxygen abundances in the CGM.
An example of gas capture can be seen in the Milky Way system as the Magellanic Stream.
The total gas mass (atomic plus ionised) of the Magellanic Stream is $\sim 10^{9}M_{\sun}$  \citep{Fox2014,D'Onghia2016}.
Two principal filaments of the Magellanic Stream show different gas-phase chemical abundances. The filament connected to the Large Magellanic Cloud shows LMC-like abundances ($\sim 0.5$ solar,
12 + log(O/H) $\sim 8.4$), and the other filament shows the Small Magellanic Cloud-like abundances, $\sim 0.1-0.2$ solar \citep{Fox2013,Richter2013,D'Onghia2016}. 
Thus, the interaction with Magellanic Cloud-like satellites can explain the values of the oxygen abundance beyond the outer break radii obtained in the MaNGA galaxies of our sample. 

The spots outside the optical radius in the surface brightness distributions in some our galaxies (e.g. M-12094-12705, panel (a1) in Fig.~\ref{figure:sequence};
M-8595-12702, panel (a2) in Fig.~\ref{figure:sequence}; M-11946-12702, panel (a3) in Fig.~\ref{figure:sequence}) can be attributed to the fragments (debris) of the captured
and destroyed dwarf galaxy, that is, the star-forming circumgalactic medium can contain not only the gas from the captured companion galaxy, but also star fragments. This justifies
the assumption that the interaction with (capture of) a small companion causes the flat gradient (mixing) in the circumgalactic medium in these galaxies. 

An examination of the morphology is the most common way to identify galaxy mergers or interactions. The asymmetry parameter, $A$, quantifies the  asymmetry of the light distribution across the galaxy
\citep{Schade1995,Conselice2003}.  The difference between the flux of a spaxel in the original image and the flux of the same spaxel after the image has been rotated by 180$\degr$ about the
centre of the galaxy is compared. The sum is carried out over all spixels within the optical radius of the galaxy.
The index $R_{A}$ specifies the distribution of residual fluxes after the model flux is subtracted \citep{Schade1995}.
The photometric profile of the galaxy was obtained (as described in Sect.~2.2), and the radial surface brightness distribution was fitted within the optical radius by a broken exponential
profile for the disc and by a general Sérsic profile for the bulge. The residual flux for each spaxel was estimated as the difference between the measured flux and flux obtained from
the fit relation for a galactocentric distance of the spaxel. The obtained residual fluxes were used to determine the index $R_{A}$.
The average value of the asymmetry parameters in early-type spirals (Sa and Sb galaxies) is $A = 0.07\pm 0.04$, and late-type spirals (Sc and Sd galaxies) show
higher asymmetries, $A = 0.15\pm 0.06$ \citep{Conselice2003}.  The high asymmetry value ($A > 0.35$) can serve as an indicator of a merger or interaction  \citep{Conselice2003,Wilkinson2022}.  
A galaxy is classified as asymmetric when  $R_{A} > 0.05$  \citep{Schade1995}. 

Figure~\ref{figure:a-ra} shows the $R_{A} - A$ diagram for our sample of galaxies. Inspection of Fig.~\ref{figure:a-ra}  (see also Table~\ref{table:general}) shows that 
the values of the asymmetry parameter for galaxies with prominent features in the surface brightness distributions (M-12094-12705, M-8595-12702, and M-11946-12702) are higher than
the average value for spiral galaxies, but they are below the threshold value of $A = 0.35$ for the merger or interaction \citep{Conselice2003,Wilkinson2022}. This is expected, because
first, the asymmetry parameter is determined for the image within the optical radius, while the features in the  surface brightness distributions (debris of the captured companion)
are outside the optical radius (Fig.~\ref{figure:sequence}). Second, the significant asymmetry in the light distribution might be produced by a 
strong interaction or merger. Strong interactions can also result in a  significant distortion of the line-of-sight velocity field, however. These galaxies were excluded from our consideration
according to our selection criteria. The line-of-sight velocity fields in our galaxies are not perfectly regular, but the distortion is weak (panels (b1, b2, b3)
in Fig.~\ref{figure:sequence}). Thus, the value of the asymmetry parameter $A$ below the threshold value  for the merger or interaction does not exclude the interaction with (capture of)
a small companion during the current epoch or in the recent past.  Therefore, we assumed that the interaction with (capture of) a small companion
causes the flat gradient (mixing) in the circumgalactic medium in all our galaxies.  
The values of the asymmetry indicator of the distribution of the residual spaxel flux $R_{A}$ exceed the threshold value of $R_{A} = 0.05$ which divides the symmetric
and asymmetric galaxies \citep{Schade1995}, that is, our galaxies should be classified as asymmetric galaxies.

The locations of our galaxies  in the $A - R_{A}$ diagram form a monotonic sequence (Fig.~\ref{figure:a-ra}).  
Galaxy merger (interaction) features may persist for up to $\sim 1$~Gyr, but they gradually fade and become faint at $\sim 200$~Myr after coalescence \citep{Lotz2008,Pawlik2016,Wilkinson2022}.
The decrease in the asymmetry parameter can be attributed either to the decrease in the mass of the captured companion or to the increase in time after the capture.
If our assumption that the interaction with (capture of) a small satellite causes the mixing of the circumgalactic medium is correct, then the flat gradient in the circumgalactic
medium is a long-living sign of the capture of a small companion.

\subsection{Origin of the inner break in the metallicity gradient}

\begin{figure}
\begin{center}
\resizebox{1.00\hsize}{!}{\includegraphics[angle=000]{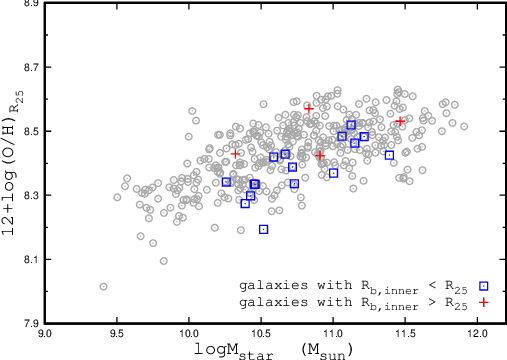}}
\caption{Oxygen abundance at the optical radius $R_{25}$ as a function of the stellar mass.
  The squares designate the galaxies of our sample in which the radial position of the inner break $R_{\rm b,inner}$ is within the optical radius $R_{25}$.
  The plus signs mark the galaxies in which the radial position of the inner break is outside the optical radius.
  The grey circles show the comparison galaxies. 
}
\label{figure:m-oh25}
\end{center}
\end{figure}

The galaxies of our sample with the flat abundance gradient in the circumgalactic medium beyond the outer breaks also demonstrate another distinctive characteristic common to all these galaxies:
They show the inner breaks in the radial oxygen abundance distributions (Fig.~\ref{figure:r-oh-manga}).
\citet{SanchezMenguiano2018} and \citet{Cardoso2025} also noted that galaxies can simultaneously show an outer flattening and  an inner drop in the oxygen abundances.
A close inspection of Fig.~\ref{figure:r-oh-manga} shows that the spaxels of the
low (CGM-like) metallicities can be found in the outer parts of the discs alone, beyond the inner break radii. This suggests that either the low-metallicity gas (captured from
the companion galaxy) infalls onto the outer part of the galaxy alone or that the low-metallicity gas inflows from the CGM as the radial gas flow. 

The classical models of inside-out growth \citep[e.g.][]{Matteucci1989} reproduce many chemical galaxy properties. Radial gas flows with radial velocities
of the order of a few km\,s$^{-1}$ are thought to take place and can play an appreciable role in the chemical evolution of galaxies, in particular, in the development of abundance gradients across 
the disc \citep[][among many others]{LaceyFall1985,Portinari2000,Mott2013,Lyu2025}.
The gas-phase metallicity in the annulus at a given radius is regulated by the interplay between the inflow, the star formation, and the outflow.  

Several works were devoted to direct measurements of radial gas velocities in galaxies. 
\citet{Wong2004} examined the CO and H\,{\sc i} velocity fields of 7 nearby spiral galaxies.  No unambiguous evidence for radial inflows was found in any of the seven galaxies,
and they obtained an upper limit of $\sim5-10$~km~s$^{-1}$ on the magnitude of any radial inflow in the inner regions of NGC~4414, NGC~5033, and NGC~5055.
\citet{Schmidt2016} investigated the radial gas flows primarily in the outer parts (up to 3~$R_{25}$) of 10 nearby spiral galaxies using high-resolution Very Large Array data from
the H\,{\sc i} Nearby Galaxy Survey (THINGS). 
They found clear indications of radial gas flows for NGC~2403 and NGC~3198 and to a lesser degree for NGC~7331, NGC~2903, and NGC~6946. The mass flow rates are of the same order, but usually higher
than the star formation rates. 
\citet{DiTeodoro2021} determined radial velocities and mass flow rates in a sample of 54 local spiral galaxies. They found that most galaxies show radial flows of only a few
km~s$^{-1}$ throughout their discs, either inward or outward. The gas mass flow rates for most galaxies are lower than 1~$M_{\sun}$~yr$^{-1}$, which means an average inflow rate of 0.3~$M_{\sun}$~yr$^{-1}$ 
outside the optical disc. These inflow rates are lower than the average star formation rate of 1.4~$M_{\sun}$~yr$^{-1}$. 
Thus, the available measurements of the radial gas velocities in galaxies are controversial and cannot provide a solid argument for or against the radial gas motion in galaxies. 

We adopted the following evolution scenario for the radial abundance profile within a framework of the model for the chemical evolution of galaxies with a  radial flow.
In the absence of the interaction (merger), a galaxy is in equilibrium condition, where the gas-phase metallicity in the annulus at a given radius is regulated by an interplay between an inflow,
the star formation, and an outflow. The radial abundance distribution throughout the whole disc showed the gradient measured in the inner disc.
When the gas captured from the satellite is mixed into the circumgalactic medium,  then the surface mass density of the CGM increases and its metallicity can also be altered. 
The inflow into the CGM of a higher density (change in the gas inflow rate) destroys the previous equilibrium condition, and a new equilibrium condition appears with time. The inner
break radius marks the boundary between the regions of old and new equilibrium conditions. The radial size of the outer zone (difference between  $R_{\rm b,outer}$ and $R_{\rm b,inner}$)
can be considered as some indicator of the time after the change in the gas inflow rate. Thus, we assumed that 
the interaction or merger increases the density of the CGM, and consequently, changes the inflow rate at the outer edge of the galaxy. The boundary between old and new inflow rates
moves towards the centre of the galaxy. The inner break in the metallicity gradient might be related to this boundary. We do not pretend that we constructed
the model for the interaction or capture. Our oversimplified scenario helped us to interpret our observational data. 
  
We considered our galaxies from this point of view. 
In galaxies with a small radial outer zone (e.g. M-11945-12705, panel (b5) in Fig.~\ref{figure:r-oh-manga}; M-8262-12704, panel (a2) in  Fig.~\ref{figure:r-oh-manga}),
the outer part of the galaxy disc between $R_{\rm b,inner}$ and  $R_{\rm b,outer}$ involves the spaxels of the CGM-like metallicities and the spaxels of metallicities corresponding
to the extrapolation of the inner metallicity gradient.  This shows that no new equilibrium reached (and no azimuthal mixing occurred) at the edge annulus in these galaxies
because the time after the change in the gas inflow rate was too short.
The radial size of the outer zone in M-12094-12705 (panel (d5) in Fig.~\ref{figure:r-oh-manga}) is very small. A discontinuity in the radial distribution
of the oxygen abundance takes place, where the radial abundance gradient turns from exponential to flat. \citet{Bresolin2009} reported a radial abundance profile like this in the nearby galaxy NGC~5236
(see panel (d) of  Fig.~\ref{figure:r-oh-nearby} here). The above examination of the morphologies and velocity fields for our sample galaxies (Fig.~\ref{figure:sequence}) 
suggests that the interaction of M-12094-12705 with a satellite occurs in an earlier stage than in other galaxies, that is, 
the galaxy M-12094-12705 might correspond to the beginning stage of the change in the gas infall rate. 
\citet{Bresolin2009} noted that peculiarities in the chemical abundance distribution in NGC~5236 might arise from the gravitational interaction with one or
more dwarf galaxies about 1~Gyr ago. 
In galaxies with large radial size of the outer zone (e.g. M-8443-12703, panel (b2) in Fig.~\ref{figure:r-oh-manga}; M-9485-12705, panel (d3) in  Fig.~\ref{figure:r-oh-manga}; 
M-11946-12702, panel (c5) in  Fig.~\ref{figure:r-oh-manga}), the abundance of the gas in the edge annulus of the galaxy is close to the metallicity of the inflowing CGM gas
(no metallicity spaxels correspond to the extrapolation of the inner metallicity gradient), that is, a new equilibrium condition already appeared.

An increase in the gas inflow rate results in steepening of the metallicity gradient in the outer disc, and consequently, in a decrease in the oxygen abundance there. 
Fig.~\ref{figure:m-oh25} shows the oxygen abundance at the optical radius $R_{25}$ as a function of the stellar mass.
The positions of the galaxies of our sample in which the radial position of the inner break $R_{\rm b,inner}$ is within the optical radius $R_{25}$
are shifted  towards the lower envelope of the band occupied by the galaxies in the (O/H)$_{\rm R_{25}}$ -- $M_{\star}$ diagram on average. 

We emphasise that the inner break of the oxygen abundance distribution across the disc can be caused not only by the change in the gas inflow rate.  
In our previous paper, we have examined a sample of the MaNGA galaxies in which the metallicity in the inner region of the disc is at a nearly constant level and the gradient
is negative at larger radii, that is, galaxies with level-slope (LS) gradients \citep{Pilyugin2024}. It has been shown that the observed behaviour of the oxygen abundances with
radius in these galaxies can be explained by the variation in the star formation history along the radius. The high oxygen abundance and the lack of systematic
variation with radius in the inner galaxy zone imply that the region has reached a high astration level (a small gas mas fraction) and the star formation rate in the region is reduced.
The observed behaviour of the oxygen abundances with radius clearly shows the effect of the inside-out disc evolution model; the galactic centre evolves more rapidly than regions
at greater galactocentric distances \citep{Matteucci1989, Pagel1995, Bergemann2014}.

In some cases (but not in any case),  the metallicity break caused by the change in the gas inflow rate can be distinguished from the break caused by the exhausting of the gas in the
central zone of the galaxy.  In the latter case, the metallicity in the inner region of the disc is at a nearly constant level and the gradient is negative at larger radii; the level-slope
(LS) gradient.  In the former case, the metallicity gradient in the inner region of the disc can be as close to zero (LS gradient) as the negative (slope-slope, SS) gradient, depending on
the evolutionary stage of the galaxy. Thus, if the galaxy shows an SS gradient, then the break in the radial abundance distribution should be attributed to the change in the gas
inflow rate.  If the galaxy shows an LS gradient, then both mechanisms might cause the break (e.g. M-10844-12705, panel (a5) in Fig.~\ref{figure:r-oh-manga};
M-10946-12702, panel (c5) in Fig.~\ref{figure:r-oh-manga}).

\section{Conclusions}

We considered the  abundance distributions in 20 massive (log($M_{\star}/M_{\sun}$) $\ga$ 10) spiral galaxies from the MaNGA survey in which the radial abundance gradients are flat 
(oxygen abundances are at a nearly constant level) at large radii, beyond the outer break radius, $R_{\rm b,outer}$. The oxygen abundances were estimated in the spaxels where the spectra
are H\,{\sc ii} region-like according to both the BPT and the WHaD classification diagrams. We examined radial abundance distributions using the median values
of abundances in bins of 0.05~dex in $R/R_{25}$. We selected the MaNGA galaxies in which the abundances beyond the outer break radius were estimated in at least five bins.

We assumed that the outer break radius, $R_{\rm b,outer}$,  divides the galaxy disc and the circumgalactic medium (CGM). The values of $R_{\rm b,outer}$ range from
$\sim 0.8\,R_{25}$ to $\sim 1.45\,R_{25}$, where $R_{25}$ is the optical (isophotal) radius of the galaxy.
The radial position of the outer break does not correlate with the stellar mass. The mean value for our sample is  $R_{\rm b,outer}/R_{25} = 1.08$, with a scatter of 0.14.
The oxygen abundances beyond the outer break radius range from 12+log(O/H) $\sim 8.0$ to $\sim 8.5$ and tend to increase with stellar mass.

The oxygen abundance distribution in each investigated galaxy also shows the inner break in the radial abundance profile at a radius $R_{\rm b,inner}$. The inner break radius divides the inner
and outer parts of the disc with different gradients. The metallicity gradient in the outer part of the galaxy ($R_{\rm b,inner} < R < R_{\rm b,outer}$) is steeper than in the inner
part ($R < R_{\rm b,inner}$). The values of $R_{\rm b,inner}$ for our sample of galaxies range from $\sim 0.45$ to $\sim 1.15$ of the optical radius $R_{25}$. 
There is no appreciable correlation between the  radial position of the inner break and the stellar mass. 

The behaviour of the radial abundance distributions in these galaxies can be explained by assuming an interaction with (capture of the gas from) a small companion and adopting
the model for the chemical evolution of galaxies with a radial gas flow. The interaction with the companion results in the mixing of the gas and in the flat metallicity gradient in the CGM.
The capture of the gas from the companion increases the radial gas inflow rate and changes the slope of the radial abundance gradient in the outer part of the galaxy.

\begin{acknowledgements}
We are grateful to the referee for his/her constructive comments. \\ 
L.S.P acknowledges support from the Research Council of Lithuania (LMTLT) (grant No. P-LU-PAR-23-28). \\
This research has made use of the NASA/IPAC Extragalactic Database (NED), which
is funded by the National Aeronautics and Space Administration and operated by
the California Institute of Technology.  \\
We acknowledge the usage of the HyperLeda database (http://leda.univ-lyon1.fr). \\
Funding for SDSS-III has been provided by the Alfred P. Sloan Foundation,
the Participating Institutions, the National Science Foundation,
and the U.S. Department of Energy Office of Science.
The SDSS-III web site is http://www.sdss3.org/. \\
Funding for the Sloan Digital Sky Survey IV has been provided by the
Alfred P. Sloan Foundation, the U.S. Department of Energy Office of Science,
and the Participating Institutions. SDSS-IV acknowledges
support and resources from the Center for High-Performance Computing at
the University of Utah. The SDSS web site is www.sdss.org. \\
SDSS-IV is managed by the Astrophysical Research Consortium for the 
Participating Institutions of the SDSS Collaboration including the 
Brazilian Participation Group, the Carnegie Institution for Science, 
Carnegie Mellon University, the Chilean Participation Group,
the French Participation Group, Harvard-Smithsonian Center for Astrophysics, 
Instituto de Astrof\'isica de Canarias, The Johns Hopkins University, 
Kavli Institute for the Physics and Mathematics of the Universe (IPMU) / 
University of Tokyo, Lawrence Berkeley National Laboratory, 
Leibniz Institut f\"ur Astrophysik Potsdam (AIP),  
Max-Planck-Institut f\"ur Astronomie (MPIA Heidelberg), 
Max-Planck-Institut f\"ur Astrophysik (MPA Garching), 
Max-Planck-Institut f\"ur Extraterrestrische Physik (MPE), 
National Astronomical Observatories of China, New Mexico State University, 
New York University, University of Notre Dame, 
Observat\'ario Nacional / MCTI, The Ohio State University, 
Pennsylvania State University, Shanghai Astronomical Observatory, 
United Kingdom Participation Group,
Universidad Nacional Aut\'onoma de M\'exico, University of Arizona, 
University of Colorado Boulder, University of Oxford, University of Portsmouth, 
University of Utah, University of Virginia, University of Washington, University of Wisconsin, 
Vanderbilt University, and Yale University.
\end{acknowledgements}

\end{document}